\documentstyle[aps,psfig]{revtex}
\twocolumn

\begin{document}
\title{Reconstruction of the spin state }
\author{Z. Hradil{\cite{optika}}, J. Summhammer, G. Badurek, H. Rauch}
\address{Atominstitut der \"{O}sterreichischen Universit\"{a}ten,\\
Stadionallee 2,\\
A--1020 Wien, Austria}
\date{\today}
\maketitle

\begin{abstract}
System of  $1/2$ spin particles  is observed repeatedly using
Stern--Gerlach apparatuses with rotated orientations. Synthesis of
such non--commuting observables  is analyzed using maximum
likelihood estimation as an example of quantum state
reconstruction. Repeated  incompatible observations represent a new
generalized  measurement. This idealized  scheme will serve for
analysis of future  experiments in neutron and quantum optics.
\end{abstract}


\section{ Introduction}

Quantum mechanics of  $1/2 $ spin particles serves often  as an
illustrating example   for many   quantum considerations in
standard textbooks of quantum theory \cite{Sakurai}. The importance
of this model is enhanced  by the fact that such states represent
the  smallest amount of  quantum  information -- quantum bits
(q-bits). Besides theoretically valuable ``Gedanken" experiments,
the spin 1/2 particles such as electrons, neutrons or polarization
states of light quanta are convenient for feasible experiments in
matter wave and quantum  optics.
 They play  crucial role in many sophisticated schemes  involving
 entanglement,   Bell state analysis or  teleportation. Several
approaches for measurement and  estimation of spin states have been
considered  recently \cite{DukesLarson,Jones,St,Buzek}. In this
Brief Report, the maximum likelihood  (MaxLik) estimation of 1/2
spin state will be formulated as an illustrating example of
 more general treatment \cite{hradil,HSR}. The given formulation
shows a tight relation between quantum and statistical theories.
Synthesis of many independent and nonequivalent ideal detections of
 Stern--Gerlach  (SG) type will be interpreted as a new generalized
measurement, output of which the quantum state is. This
consideration will be used as a basic tool for further
investigation in depolarization measurements, neutron and light
interferometry and  quantum state reconstruction.

Basic properties of spin 1/2 quantum systems will be briefly
reviewed.  A pure state  (projector) may be represented by the
expression
\begin{equation}
|{\bf a} \rangle \langle {\bf a} | = \frac{1}{2}(1 + a_i \sigma_i) ,
\label{project}
\end{equation}
where ${\bf a} = (a_1,a_2,a_3)$ is the three-dimensional normalized
vector, $\sigma_i, i = 1,2,3 $ represent the Pauli matrices and the
summation convention for repeated indices is used. Since
\[
\sigma_i \sigma_j = \delta_{ij} + i \epsilon_{ijk}\sigma_k,
\]
the scalar product of two projectors is given as
\[
|\langle a|b\rangle|^2 = \frac{1}{2}(1 + a_i b_i).
\]
General state described by a density matrix may be parameterized by
\begin{eqnarray}
\hat \rho = p_+ |{\bf a} \rangle \langle {\bf a} | + p_- |{\bf - a} 
\rangle
\langle {\bf - a} | \\
\frac{1}{2} + \frac{1}{2} \sigma_i a_i (p_+ - p_-),
\end{eqnarray}
where $p_+ + p_- = 1 $ and the states $|{\bf a} \rangle, |{\bf - a}
\rangle$ denote a general orthogonal basis.
Spin  state may be alternatively  described by a polarization
vector
\begin{equation}
r_i =   \langle \sigma_i \rangle  = a_i (p_+ - p_-),
\end{equation}
where the brackets $\langle \rangle $ denote an expectation value.
Hence  the polarization  $r_i$ completely determines the state of
quantum system. Degree of polarization may be introduced as
\[
|{\bf r}|^2  \le 1
\]
and  $|{\bf r}|^2 = 0 $
for completely unpolarized (mixed) state and
$|{\bf r}|^2 = 1 $ for completely polarized  (pure) states.

The polarization or spin may be measured by
 projecting the state into the given directions of SG apparatus
 $\bf\pm  a.$ Closure relation  and operator representation of such
 a device  simply read
\begin{eqnarray}
|{\bf a }\rangle \langle{\bf a} | + |{\bf - a }\rangle
\langle{\bf - a } | = \hat 1,
\label{closure}
\\
\label{apparatus}
\hat A = \frac{1}{2}\bigl[|{\bf a }\rangle \langle{\bf a} | - |{\bf - a }\rangle
\langle{\bf - a } | \bigr].
\end{eqnarray}
 Assuming for the sake of  simplicity  always  the same total
number of particles N, the number of particles with the spin ``up"
and ``down" estimates  projections of the polarization vector
according to the relations
\begin{equation}
n_{\pm} = N p( {\bf \pm a}) = \frac{1}{2} N ( 1 \pm {\bf r a}).
\end{equation}
Since this may be done  for three  orthogonal directions of
coordinate axes  ${\bf x_i}, i = 1,2,3, $ the polarization  may be
found by eliminating the
 total number of particles $N $
\begin{equation}
r_i = \frac{n_{i+} - n_{i-}} {n_{i+} + n_{i-}}.
\label{det}
\end{equation}
Each polarization component  is  determined separately. This
represent a correct  solution, provided that resulting polarization
is inside the Poincar\'{e}
 sphere  $|{\bf r}|^2 \le 1 $ only.  This example  has been used
for motivation of general analysis of quantum state reconstruction
in Ref. \cite{HSR}. As demonstrated,  the ``states" outside the
Poincar\'{e} sphere violate the  positive semidefiniteness  of
quantum states yielding improper  quantum description  of  noises.
Similar problems appear in the case  when more than three
projections  are used.  Some results of SG projections  might
appear as incompatible among themselves  due to the fluctuations
and noises  involved. Various SG  measurements are  not equivalent,
since they are observing various ``faces" of the spin system. Such
measurements, even when done with equal number of particles,
determine various  projection with different errors.
  Detected data $ n_{i,{\pm}}$ collected from SG observations in M
various directions ${\pm \bf a}^i$, $i =1,2, \ldots M $
 sample various  binomial distributions. Significantly, the
 detected data $n_{i,\pm}$   fluctuate with the  root--mean square
errors  given by   $\sqrt{N (1- ({\bf r}{\bf a}^j)^2) }/2$
  depending on the deviations between projections and the true (but
unknown!) direction of the spin {\bf r}. Various projections cannot
be therefore trusted with the same degree of credibility, since
they are affected by different errors. This is manifested in
quantum theory, since various SG measurements  are incompatible and
corresponding  operators (\ref{project}) do not commute  for
different orientations ${\bf a}^j$. Such data cannot be obtained in
the same measurement, but may be
 collected  by repeating. An optimal   procedure must predict an
unknown state and simultaneously take into account data
fluctuations. This indicates the nonlinearities of an  algorithm.
MaxLik estimation does this job and fits the data to a quantum
state.  Besides this, it is the only procedure, which provides the
same structure as generalized measurement \cite{new}.
 Henceforth,  synthesis of incompatible measurements may be
interpreted as a novel measurement of quantum state. This will be
demonstrated in the following section.

\section{Spin estimation}

Provided that source supplies $1/2$ spin  particles prepared in the
same mixed state, an ideal lossless  SG measurement  performed
repeatedly on the system of   $N$  particles will be assumed. The
setting of SG apparatus may change. Provided that detection has
been done with $M $ different settings, $N*M $ particles have been
used altogether and an unknown quantum state should be found.
  The results of the measurement may be characterized by settings
of the SG apparatus $ {\pm \bf a}^j$ and by the relative
frequencies of the outcomes  $1/2(1\pm X_j) = n_{j,\pm}/N.$ The
question is what state(s) fit(s) the data in optimal way. One might
be tempted to sample and invert the probability, predicted by
quantum theory, as it is done in the case of equation (\ref{det}).
Because each SG detection is represented by a complete measurement,
the  sum  the relations (\ref{closure}) for each setting of SG
apparatus $j$ reads
\begin{eqnarray}
\frac{1}{M}\sum_j^M
|{ \bf a}^j \rangle \langle { \bf a}^j | + |{- \bf a}^j \rangle
\langle {- \bf a}^j | = \hat 1.
\label{id2}
\end{eqnarray}
  However, the  expected relations
\begin{eqnarray}
{\rm Tr} \{ \hat  \rho \frac{1}{M}
|{\pm \bf a}^j \rangle \langle {\pm \bf a}^j | \} = \frac{1}{2M}(1 \pm X_j).
\label{expect2}
\end{eqnarray}
cannot be fulfilled, in general, since the system is overcompleted
and data are fluctuating. Hence, the probabilities cannot be mapped
so straightforwardly with the   relative frequencies of outcomes.

 MaxLik principle provides a tool, how to treat this problem.
 The most probable state  consistent with the data should be found.
As the measure of probability, the likelihood functional
corresponding to the product of all the probabilities for  all
detected data  may be constructed
\begin{equation}
{\cal L}({\hat \rho}) =
\prod_{j}  \biggl(\langle {\bf a}^j | \hat \rho| {\bf a}^j \rangle
\biggr)^{N(1+ X_j)/2}
\biggl(\langle {-\bf a}^j | \hat \rho| {-\bf a}^j \rangle
\biggr)^{N(1- X_j)/2}     .
\end{equation}
Extremal  states of likelihood functional satisfy the nonlinear
operator equation \cite{HSR}
\begin{equation}
\frac{1}{2 M} \sum_j \biggl[(1+ X_j) \frac{|{\bf a}^j\rangle \langle
{\bf a}^j |} { \langle {\bf a}^j| \hat \rho |{\bf a}^j \rangle }  +
(1- X_j)\frac{  |{\bf- a}^j \rangle \langle {\bf- a}^j |}
{ \langle {\bf- a}^j|\hat \rho |{\bf-a}^j \rangle }
 \biggr] \hat \rho= \hat  \rho
 \label{ex}
\end{equation}
Quantum state may be represented by polarization.  Using the
relation   (\ref{project}), multiplying both the sides by
${\sigma_k}$ and performing the trace, the equation reads
\begin{equation}
R({\bf r} ) {\bf r}  + {\bf K}({\bf r}) + i {\bf K}({\bf r})
\times {\bf r}  = {\bf r},
\end{equation}
where  the functions are defined as
\begin{eqnarray}
 R({\bf r}) = \frac{1}{2 M} \sum_j
\biggl( \frac{1+ X_j}{1+ {\bf  a}^j {\bf r} } +
\frac{1- X_j}{1- {\bf  a}^j {\bf r} } \biggr) ,
\nonumber
\\
{\bf K}({\bf r}) = \frac{1}{2 M} \sum_j
\biggl( \frac{1+ X_j}{1+ {\bf  a}^j {\bf r}}
- \frac{1- X_j}{1- {\bf  a}^j {\bf r} } \biggr) {\bf a}^j.
\nonumber
\end{eqnarray}
Because real and imaginary parts are dependent, the real part of
this equation  represents  sufficient and necessary conditions.
The final equation for polarization  reads
\begin{equation}
R({\bf r} ) {\bf r}  + {\bf K}({\bf r})  = {\bf r}.
\label{iter}
\end{equation}
The relation  (\ref{iter}) may be    advantageously used for
iteration of the  solution. Starting from the centre of
Poincar\'{e} sphere ${\bf r} = 0,$ the left hand side of  eq.
(\ref{iter})  provides the first correction, which may be used for
subsequent iteration, etc.. This procedure provides a quick
algorithm for MaxLik fitting of an unknown quantum state inside the
Poincar\'{e} sphere.

An equivalent  result may be derived, provided that likelihood
function is parameterized directly using the polarization. The
relevant part of  the
 likelihood function corresponding to the observation of particular
data   reads
\begin{equation}
{\cal L}(({\bf r})) =
\prod_{j} (1 + {\bf r}{\bf a}^j)^{N(1+ X_j)/2}
(1 - {\bf r}{\bf a}^j)^{N(1- X_j)/2}  .
\label{ideallik}
\end{equation}
The vector ${\bf r}$  parameterizes an unknown polarization inside
the Poincar\'{e} sphere and the products runs over all M
directions. The standard statistical approach using MaxLik
    $\frac{\partial}{\partial \bf r}  \ln{\cal L}$ provides the
vector  equation  for extremal polarization   \cite{Jones}
\begin{equation}
\sum_j \frac{X_j -{\bf a}^j {\bf r}}{1 - ({\bf a}^j {\bf r})^2} {\bf
a}^j =
0.
\label{standard}
\end{equation}
The equation (\ref{iter}) is equivalent to the
equation (\ref{standard}). Indeed, the  equation
 (\ref{standard})  is nothing else as
 $ {\bf K}({\bf r}) = 0,$ implying
the relation $R({\bf r} ) = 1.$  On the other hand, the equation
(\ref{iter})  may be rewritten to the form of equation (\ref{standard}) as well.

\begin{figure}
 \centerline{\hspace{1cm}\psfig{file=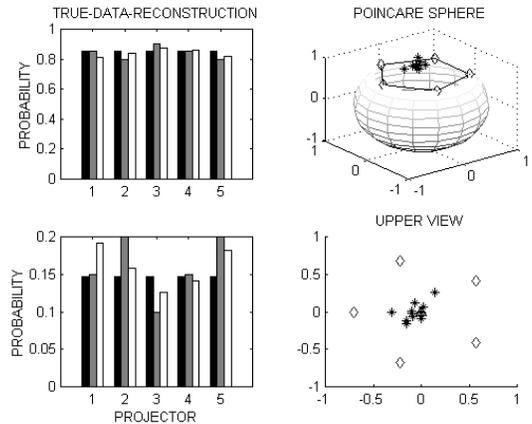,width=10.5 cm,clip=}}
 \vspace{-9cm}
\caption{Results and interpretation of  spin reconstruction for
 numerical simulation  of Stern--Gerlach detection with apparatuses
 in five various settings}
\label{Fig1}
\end{figure}

Results  of numerical simulation are shown in the Fig. \ref{Fig1}.
Stern-Gerlach detection is simulated here  for projection of an
unknown state (north pole on Poincar\'{e} sphere) in five various
directions. Each measurement is done with 20 impinging particles,
which are registered either with the spin up (upper left panel)
either down (lower left panel). Both the left panels show typical
values  for a single experiment. For each position of the projector
there are three bars in the upper and lower left panels. The first
bars (black) show the true value of the probability. The second
 bars (gray) show the counted statistics fluctuating around the
 true value of probability. The third bars (hollow) show results of
 the  reconstruction -- the statistics of reconstructed state
 corresponding to the given projector. Notice here, that upper and
 lower panels are complementary and sum of corresponding
 probabilities is one.
  The  right panels visualize  results of 10 times repeated
 experiment on the Poincar\'{e} sphere. Symbols of diamonds  denote
 the positions of five projectors. Orthogonal projectors in
 opposite directions are not depicted. The stars show the position
 of  reconstructed states. The true state corresponds to the north
 pole. Lower right panel show the upper view.

Quantum formulation  posses a nontrivial interpretation, which can
hardly be recognized in the equation for polarization
(\ref{standard}). Because the above   scheme  determines a quantum
state, it must  exist a generalized measurement described  by a
probability operator measure (POM) \cite{Hel76}, result of which
the quantum state is. Really, such a   probability operator measure
can be easily  find by   proper renormalization of original  SG
measurement \cite{HSR}. Let us define the  POM  as renormalized SG
projectors
\begin{eqnarray}
|{\pm \bf a}^j \rangle \langle {\pm \bf a}^j |_{R} =
\frac{1 \pm X_j}{2M \langle{\pm \bf a}^j|\hat \rho_e|{\pm \bf a}^j)} 
|
{\pm \bf a}^j \rangle \langle{\pm \bf a}^j | ,
\end{eqnarray}
for each index $j$.
 The closure relation  then reads
\begin{eqnarray}
\sum_j^M
|{ \bf a}^j \rangle \langle { \bf a}^j |_{R} + |{- \bf a}^j \rangle
\langle {- \bf a}^j |_{R} = \hat 1,
\label{id}
\end{eqnarray}
and the renormalized POM fulfills the conditions
\begin{eqnarray}
{\rm Tr} \{ \hat  \rho_{est}
|{\pm \bf a}^j \rangle \langle {\pm \bf a}^j |_{R} \} = \frac{1}{2M}(1 \pm X_j).
\label{expect}
\end{eqnarray}
Here  $\hat \rho_e$
denotes the  extremal state -- a solution of the  equation (\ref{iter}).
Indeed, the relation (\ref{id}) coincides with the equation for
extremal states (\ref{ex}),
whereas the condition for expectation values (\ref{expect})
is fulfilled as an  identity.
 The reconstruction is done  on the subspace, where the
renormalized POM reproduces  the  identity operator. Particularly,
this means that the identity operator on the right hand side of
(\ref{id}) is spanned by the  one dimensional subspace only (i.e by
a single ray), provided that the  extremal state  $\hat \rho_{e}$
is a pure state. For a general extremal   density matrix, the
reconstruction is accomplished in the whole two dimensional Hilbert
space. The distinction between  relations (\ref{id2}),
(\ref{expect2}) and (\ref{id}), (\ref{expect}) characterize the
subtle point of quantum state reconstruction.
 MaxLik solution may be  also interpreted in the language of
probabilities. The detected data $n_{i,\pm}$ samples different
binomial probability distributions for $i= 1,
\ldots, M. $  MaxLik estimation  finds a common multinomial
distribution, sampling of which the data seems to be with the highest
likelihood.

The  method developed here may be compared with the existing
approaches. Jaynes maximum entropy principle (MaxEnt) \cite{Jaynes}
has  been applied to the estimation of spin 1/2/ states in Refs.
\cite{DukesLarson,Buzek}. In general, these methods are not
equivalent. The MaxLik solution seeks for the most likely solution
consistent with the data, whereas the MaxEnt  searches for the
worst solution still  consistent with the data. This may be
interpreted as different prior information in the  maximum
probability principle \cite{Frieden}.
  But this is not the only difference. External  conditions of both
the approaches differ substantially. MaxLik has been  applied to
the measurements with many projectors.  However, the same
conditions cannot be used in MaxEnt principle.
 Since there are only three free parameters necessary for
determination of an  unknown  spin state,  conditions
(\ref{expect2}) cannot be fulfilled, in  general. MaxEnt principle
cannot be used provided that there are more than three independent
conditions put on the density matrix of $1/2$ spin system.

In the papers Refs. \cite{optimal} an optimal strategy for
measuring an unknown  two-state system prepared in a mixed state is
investigated. As a result,  an optimal POM may be predicted. On the
other hand, not the measurement but the mathematical  treatment is
optimized in this paper. This seems to be reasonable from the
experimentalist's point of view since it is questionable how to do
a general  measurement described by a POM. As demonstrated, for
given measurement MaxLik estimation provides an  optimal treatment,
since it reproduces a quantum measurement. As argued in Ref.
\cite{porovnani} both the treatments provide comparable results.

\section{Summary}

Synthesis of incompatible observations   represented by various
settings of   SG apparatus has been evaluated using MaxLik
estimation. As the result it  defines a generalized measurement of
a quantum state. Synthesis of various incompatible observations is
again a  generalized measurement.  Developed formalism will be
applied in the future for investigation of various problems such
as:  spin estimation of neutrons in the depolarization experiments,
estimation of quantum state inside the interferometer or analysis
of entanglement.

 This work was supported by TMR Network ERB FMRXCT 96-0057
``Perfect Crystal Neutron Optics'' of the European Union, by the
East--West program of Austrian Academy of Science and by the grants
of Czech Ministry of Education VS 96028 and CEZ J14/98.

~\vspace{-4cm}

\end{document}